\documentclass[aps,prb,twocolumn,showpacs,preprintnumbers,amsmath,amssymb]{revtex4}

\usepackage{graphicx}
\usepackage{dcolumn}
\usepackage{bm}
\usepackage{amssymb}
\usepackage{graphicx}
\usepackage{amsmath}
\usepackage{xspace}

\begin{document}

\title{Short-Range B-site Ordering in Inverse Spinel Ferrite NiFe$_2$O$_4$}
\author{V.~G.~Ivanov$^1$, M.~V.~Abrashev$^1$, M. N. Iliev$^2$, M.~M.~Gospodinov$^3$, J. Meen$^2$, M.~I.~Aroyo$^4$}
\affiliation{$^1$Faculty of Physics, University of Sofia, 1164 Sofia, Bulgaria\\
$^2$Texas Center for Superconductivity and Department of Physics, University of Houston, Texas 77204-5002, USA\\
$^3$Institute of Solid State Physics, Bulgarian Academy of Sciences, 1184~Sofia, Bulgaria\\
$^4$Departamento de F\'{\i}sica de la Materia Condensada, Universidad del País Vasco, 48080 Bilbao, Spain
}

\date{\today}

\begin{abstract}
The Raman spectra of single crystals of NiFe$_2$O$_4$ were studied in various scattering configurations in close comparison with the corresponding spectra of Ni$_{0.7}$Zn$_{0.3}$Fe$_2$O$_4$ and  Fe$_3$O$_4$. The number of experimentally observed Raman modes exceeds significantly that expected for a normal spinel structure and the polarization properties of most of the Raman lines provide evidence for a microscopic symmetry lower than that given by the $Fd\bar3m$ space group. We argue that the experimental results can be explained by considering the short range 1:1 ordering of Ni$^{2+}$ and Fe$^{3+}$ at the B-sites of inverse spinel structure, most probably of tetragonal $P4_122$/$P4_322$ symmetry.
\end{abstract}

\maketitle

\section{Introduction}
The spinel ferrites with general formula AFe$_2$O$_4$ have interesting physical properties and are of  technological importance.\cite{brabers1995} In particular, NiFe$_2$O$_4$ is of increased interest as this material, in the form of bulk, powder, thin film or nanoparticles, finds or promises numerous applications in magnetic storage systems,\cite{han1996} magnetic resonance imaging,\cite{cunningham2005}, spintronics,\cite{seneor1999,hu2002} etc. At present it is accepted that NiFe$_2$O$_4$ crystallizes with inverse spinel structure,\cite{hastings1953, gorter1954,subramanyam1971,carta2009} described by the face-centered cubic (FCC) space group $Fd \bar3 m$(No.227, Z=8). In this structure the tetrahedral A-sites (8a) are occupied by half of the Fe$^{3+}$ cations, whereas the rest of the Fe$^{3+}$ and Ni$^{2+}$ cations are distributed over the octahedral B-sites (16d). A fundamental question however arises whether Fe$^{3+}$ and Ni$^{2+}$ are spread in a random fashion among the B-sites or exhibit specific short-range order on a spatial scale, which is below the detection limit of the standard diffraction techniques.  This issue can be addressed effectively by polarization Raman spectroscopy since the number, the frequencies and the polarization selection rules of the Raman-active vibrational modes are highly sensitive to the atomic short-range order. For the normal spinel structure only five Raman allowed phonons ($A_{1g} + E_g +3F_{2g}$) are expected and this is the case for number of materials such as Co$_3$O$_4$,\cite{hadjiev1988} CdCr$_2$Se$_4$,\cite{iliev1978} and Fe$_3$O$_4$ above the Verwey transition temperature, \cite{shebanova2002} to mention a few. Another group of spinels, such as NiFe$_2$O$_4$,\cite{graves1988,zhou2002} NiAl$_2$O$_4$,\cite{laguna2007} CoFe$_2$O$_4$,\cite{wang2006,ilievCFO} exhibit much richer Raman spectra with a number of additional peaks. The latter materials are with inverse or partly inverse spinel structure and it is reasonable to assume that the appearance of additional Raman lines is somehow related to the presence of non-equivalent atoms at the B-sites, which may have the following consequences for the Raman spectra:
(i) The random distribution of inequivalent B$'$ and B$''$ atoms destroys the translation symmetry, in particular of the oxygen sublattice, and activates otherwise forbidden phonon modes. One expects in this case additional broad Raman structures which  roughly reproduce the smeared one-phonon density of states;\cite{iliev2003}
(ii) A short-range ordering of B$'$ and B$''$ atoms may result in formation of domains of symmetry lower than $Fd\bar 3m$ with new sets of Raman-allowed phonons. The coexistence of twin variants of these local structures leads to a superposition of spectra corresponding to different scattering configurations.

In this work we report results of a polarized Raman study of NiFe$_2$O$_4$. The analysis is done in close comparison with lattice dynamics calculations for spinel structures with either full disorder or ordering of Ni$^{2+}$ and Fe$^{3+}$ at the B-sites as well with the corresponding spectra of Ni$_{0.7}$Zn$_{0.3}$Fe$_2$O$_4$ and Fe$_3$O$_4$. A conclusion is made that at a microscopic level the structure of NiFe$_2$O$_4$ can be considered as a mixture of twin variants of a structure with Fe$^{3+}$ and Ni$^{2+}$ ordered over the B-sites.

\section{Samples and Methods}
The first step in the growth of NiFe$_2$O$_4$ and Ni$_{0.7}$Zn$_{0.3}$Fe$_2$O$_4$ single crystals was
sintering of polycrystalline samples by solid-state reaction of stoichiometric amounts of NiO, Fe$_2$O$_3$, and ZnO
annealed for 48~h at 1150$^\circ$C in oxygen atmosphere. As a next step the high temperature solution growth method was
applied using PbO-PbF$_2$-B$_2$O$_3$ flux with ratio of the components of 0.50~:~0.48~:~0.02 for NiFe$_2$O$_4$ and of
0.67~:~0.32~:~0.01 for Ni$_{0.7}$Zn$_{0.3}$Fe$_2$O$_4$, respectively. The flux was mixed with NiFe$_2$O$_4$ powder in a 10~:~1 ratio
or with Ni$_{0.7}$Zn$_{0.3}$Fe$_2$O$_4$ in a 7~:~1 ratio and annealed in a 500~ml platinum crucible at 1225$^\circ$C in air for 48~h.
After annealing the temperature was decreased to 950$^\circ$C at a rate of 0.5$^\circ$C/h for NiFe$_2$O$_4$ and to 1000$^\circ$ at a rate of
1$^\circ$C/h for Ni$_{0.7}$Zn$_{0.3}$Fe$_2$O$_4$. The flux was decanted and the crystals of up to 5~mm in size
were removed from the bottom of the Pt crucible. These crystals were of octahedral shape with large (111) and smaller (100) and (110) facets.
The Fe$_3$O$_4$ sample was natural polycrystalline magnetite with typical grain size of 300~$\mu$m, large enough for obtaining polarized Raman spectra in exact scattering configuration from properly oriented microcrystal surfaces. The elemental content has been confirmed by x-ray wavelength dispersive spectrometry (WDS)using a JEOL JXA8600 electron microprobe analysor.

The polarized Raman spectra were measured from (100) cubic surfaces with a triple T64000 spectrometer equipped with microscope. The spectra obtained with 633~nm, 515~nm, 488~nm, or 458~nm excitation were practically the same.

\section{Results and Discussion}
\subsection{Polarized Raman spectra of NiFe$_2$O$_4$, Fe$_3$O$_4$, and Ni$_{0.7}$Zn$_{0.3}$Fe$_2$O$_4$}
From symmetry considerations one expects for the normal spinel $Fd \bar 3 m$ structure  five ($A_{1g}+E_g+3F_{2g}$) Raman active modes, which could be identified by measuring the Raman spectra in several exact scattering configurations, e.g. $XX$, $XY$, $X'X'$, and $X'Y'$. The first and second letters in these notations correspond, respectively, to the polarization of incident and scattered light, where $X$, $Y$, $X'$, and $Y'$ denote the $[100]_c$, $[010]_c$, $[110]_c$, and $[1\bar{1}0]_c$ cubic directions.  As it follows from Table~\ref{CubicSR}, one expects two Raman lines ($A_{1g}+E_g$) in the $XX$ spectrum, five lines ($A_{1g}+E_g+3F_{2g}$) in the $X'X'$ spectrum, three $(3F_{2g})$ lines in the $XY$ spectrum, and only one $(E_g)$ line in the $X'Y'$ spectrum.

In Fig.1 are compared the experimental Raman spectra of NiFe$_2$O$_4$ and the closely related spinel Fe$_3$O$_4$ obtained at room temperature with 488~nm excitation. The spectra of NiFe$_2$O$_4$ taken with 633~nm excitation are shown in more detail in Figure 2. The selection rules for the Raman bands of Fe$_3$O$_4$ follow strictly those for the normal spinel structure (see Table~\ref{CubicSR}). This is to be expected since the experimental temperature is well above the Verwey transition temperature of magnetite and, therefore, the charge is smeared uniformly among the Fe $B$-sites. In contrast to Fe$_3$O$_4$, the number of experimentally observed Raman lines in the spectra of NiFe$_2$O$_4$ exceeds significantly the number expected for a normal spinel structure and the polarization rules are strictly followed for only a few lines, namely those at 213~cm$^{-1}$($F_{2g}$), 333~cm$^{-1}$($E_g$), and 705~cm$^{-1}$($A_{1g}$). It is remarkable, however, that these spectra are practically identical to those reported earlier for NiFe$_2$O$_4$ crystals, thin films and nanocrystalline samples from different sources (see e.g. Refs.~\onlinecite{graves1988,zhou2002}). Much richer than expected Raman spectra have been also reported for other compounds, such as NiAl$_2$O$_4$, with nominally inverse spinel structures.\cite{laguna2007}
\begin{figure}[htb]
\includegraphics[width=8cm]{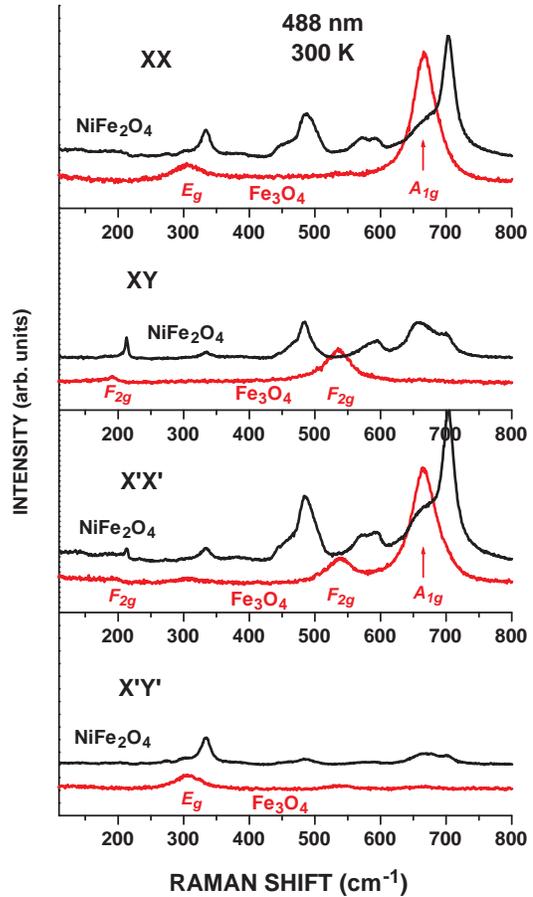}
 \caption{(Color online) Polarized Raman spectra of NiFe$_2$O$_4$ and Fe$_3$O$_4$ as obtained at room temperature with 488~nm excitation.}
\end{figure}
\begin{figure}[htb]
\includegraphics[width=8cm]{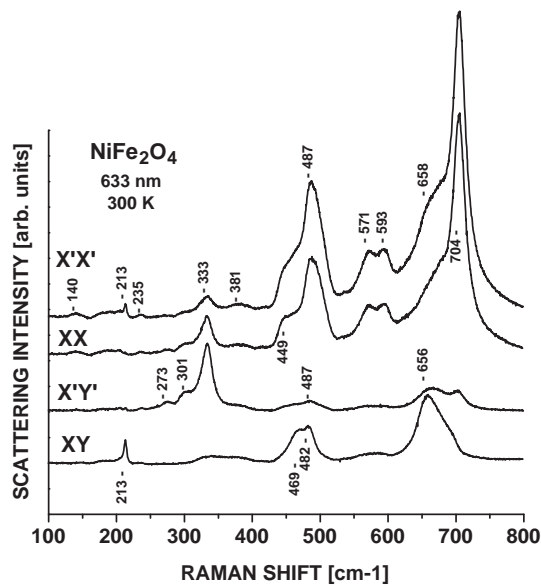}
 \caption{Polarized Raman spectra of NiFe$_2$O$_4$ as obtained at room temperature with 633~nm excitation.}
\end{figure}

The larger number of Raman active modes in inverse spinels have been discussed before and has tentatively been explained in terms of defect-induced lattice distortions due to deviation from stoichiometry and/or coexistence of 'normal' and 'inverse' domains. \cite{graves1988,laguna2007}  It seems that the deviation from the stoichiometry has little effect on the Raman spectra. In Figure 3 are compared the Raman spectra of NiFe$_2$O$_4$ and Ni$_{0.75}$Zn$_{0.25}$Fe$_2$O$_4$. The Zn$^{+2}$ ion has a larger ionic radius than Ni$^{+2}$ and is expected to increase the structural disorder of the oxygen sublattice. If the extra lines in the Raman spectra of NiFe$_2$O$_4$ were induced by a disorder, one would expect their relative intensity to increase in the Zn-substituted samples. In contrast, upon Zn substitution for Ni these lines broaden and decrease in intensity. Therefore, the additional lines in the Raman spectrum of NiFe$_2$O$_4$ indicate to a short-range order of Fe and Ni cations rather than a random distribution over the octahedral B-positions.

\begin{figure}[htb]
\includegraphics[width=8cm]{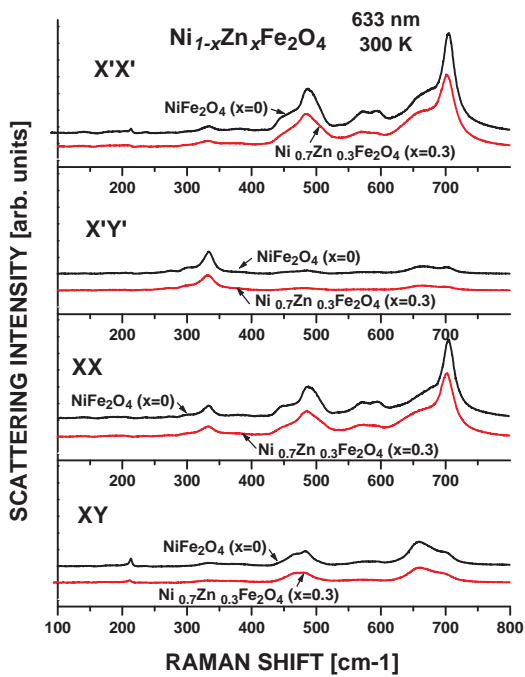}
 \caption{Polarized Raman spectra of NiFe$_2$O$_4$ and Ni$_{0.7}$Zn$_{0.3}$Fe$_2$O$_4$ as obtained at room temperature with 633~nm excitation.}
\end{figure}

\subsection{Raman Phonons in B-Site Ordered Phases in Inverse Spinels}
\subsubsection{Symmetry aspects of the 1:1 ordering at the B-sites of inverse spinel structure}

The symmetry aspects of the 1:1 ordering at the B-sites of inverse spinel structure
have been discussed in detail by Haas.\cite{haas1965} It has been shown that there are two possible types of such ordering,  $\alpha$-type and $\beta$-type, illustrated in Fig.4 and shortly described below.

The $\alpha$-type order is characterized by ...-B$''$-B$'$-B$''$-B$'$-... chains along the $[110]$ and $[1\bar{1}0]$ cubic directions. The space group is $P4_122$ (\#91) with a tetragonal unit cell two times smaller than the face-centered spinel unit cell with lattice parameters $\vec{a}_t= \frac{1}{2}(\vec{a}_c + \vec{b}_c),\  \vec{b}_t= \frac{1}{2}(\vec{a}_c-\vec{b}_c),\  \vec{c}_t=\vec{c}_c$. The same type of ordering can be described by the $P4_322$ (\#95) space group, which is enantiomorphic to $P4_122$. The atomic site symmetries and the classification of the normal modes of vibration in the two space groups are equivalent. For this reason our further analysis will be done almost exclusively in the context of $P4_122$ group.  From symmetry considerations the $Fd \bar 3 m - P4_122/P4_322$ disorder-order transition is of first order. Therefore, in the phase diagram one expects two-phase region (miscibility gap) where the cubic and the tetragonal phases coexist.  If upon cooling $\alpha$-type ordering does take place, the tetragonal axis may be aligned along each of the three equivalent $\vec{a}_c$, $\vec{b}_c$ or $\vec{c}_c$ cubic directions.  This implies that at a microscopic level six types of tetragonal domains, three types for $P4_122$ and three for $P4_322$, with mutually orthogonal 4-fold axes may coexist at room temperature (Fig.4). The enantiomorphic $P4_122/P4_322$ pairs of domains with the same orientation of the 4-fold axis will be referred by a common number, I, II and III for the cubic $X$, $Y$ and $Z$ directions respectively. In the case of domains of relatively small size ($\leq 50$ lattice constants), their  presence remains below the detection limits of standard diffraction techniques.

\begin{figure}[htb]
\includegraphics[width=8cm]{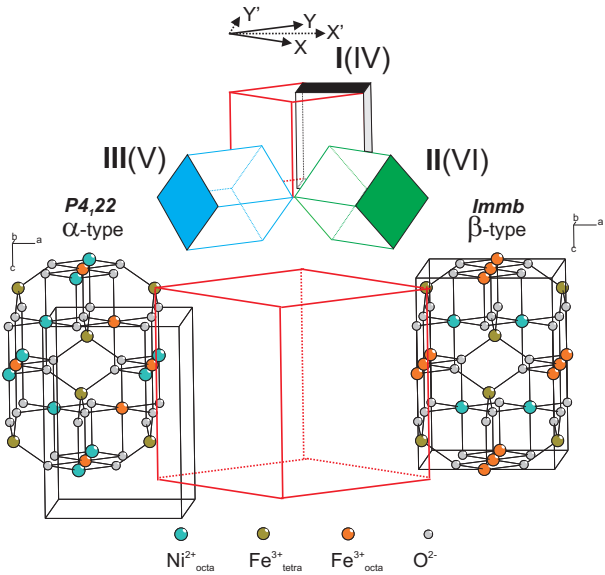}
 \caption{(Color online) $\alpha$-type ($P4_122/P4_322$) and $\beta$-type ($Imma$) ordering in inverse spinel structure. The orientation of the twin variants, I, II, III for the tetragonal structure and IV, V, VI for the orthorhombic structure, with respect to the cubic directions are also shown.}
\end{figure}

The $\beta$-type order is characterized by ...-B$'$-B$'$-B$'$-B$'$-... chains along the [110] direction and ...-B$''$-B$''$-B$''$-B$''$-... chains along the [1$\bar{1}$1] cubic directions (Fig.4). The space group is $Imma$ (\#74) with an orthorhombic unit cell two times smaller than the face-centered spinel unit cell (lattice parameters $\vec{a}_o= \frac{1}{2}(\vec{a}_c + \vec{b}_c),\  \vec{b}_o= \frac{1}{2}(\vec{a}_c-\vec{b}_c),\  \vec{c}_o=\vec{c}_c$). Here again the phase transition from spinel to orthorhombic structure is of first-order type and at a microscopic level three pairs (IV, V and VI) of mutually orthogonal domains may coexist within the framework of an averaged spinel structure.

The experimentally confirmed averaged cubic structure of NiFe$_2$O$_4$ will be compatible with tetragonal $P4_122/P4_322$ and/or orthorhombic $Imma$ structure(s) if the twin variants of these structures are uniformly oriented with respect to the cubic axes as shown in Fig.4. This means that the experimental Raman spectrum in a given cubic scattering configuration will be a superposition of spectra obtained simultaneously from tetragonal $P4_122$ twin variants in three different scattering configurations, corresponding to types (I), (II), or (III) orientation. The number of twin variants and, hence, the scattering configurations is doubled to six for the orthorhombic $Imma$ structure, accounting that $a_o$ and $b_o$ parameters are interchangeable.

Due to lower symmetry (compared to that of an ideal spinel) the number of Raman allowed modes in B-site-ordered structures increases. Their classification is given in Table~\ref{Modes}. The polarization selection rules for the B-site ordered $P4_122$ and $Imma$ structures, averaged over all twin variants, are summarized in Tables~\ref{Tetra} and~\ref{Ortho}.

The tetragonal structure gives rise to 84 normal modes (accounting for the mode degeneracy), which is twice the number of normal modes in the cubic structure. This is due to the fact that the primitive cell of $P4_122$ has two times bigger volume than the primitive cell of the FCC structure. By means of the group-subgroup relations half of the normal modes in the tetragonal structure can be mapped onto $\Gamma$-point modes of the cubic structure. For the Raman-active modes the correspondence is:
\begin{eqnarray}
\label{Tetracorr}
A_{1g} \rightarrow &A_1 \\
E_g/E_u \rightarrow &A_1+B_2\\
F_{1g}/F_{1u} \rightarrow &A_2+E\\
F_{2g}/F_{2u} \rightarrow &B_1+E
\end{eqnarray}
The rest of the $\Gamma$-point modes of the tetragonal structure originate from a zone-folding of the FCC Brillouin zone, which maps $X^*$, the star of zone-boundary $X$-point of $Fd\bar 3m$, onto $\Gamma$-point of $P4_122$. Among the Raman-active modes of the tetragonal structure such are: $5A_1+5B_1+8B_2+10E$. From a physical point of view the new spectral features in $P4_122$ can be considered as a result of splitting of the degenerated Raman modes of $Fd\bar 3m$ into doublets, as well as activation of the IR $F_{1u}$, $\Gamma$-point silent $F_{1g}$ and $F_{2u}$, and zone-boundary modes.

The primitive cell volumes of the orthorhombic $Imma$ structure and FCC structure are equal, and no zone-folding takes place. All Raman-active modes of the orthorhombic structure can be mapped onto $\Gamma$-point modes of $Fd\bar 3m$:
\begin{eqnarray}
\label{Tetracorr}
A_{1g} \rightarrow &A_g \\
E_g \rightarrow &A_g+B_{1g}\\
F_{2g} \rightarrow &A_g+B_{2g}+B_{3g}\\
F_{1g} \rightarrow &B_{1g}+B_{2g}+B_{3g}
\end{eqnarray}
Therefore, the extra Raman bands in $Imma$ should consist of a doublet originating from the $E_g$ mode, 3 triplets originating from the $F_{2g}$ modes, and a triplet resulting from activation of the silent $F_{1g}$ mode of the FCC structure.

\subsubsection{Lattice dynamics calculations of $\Gamma$-point Raman phonons  of inverse spinel NiFe$_2$O$_4$}

Theoretical results for the lattice dynamics of NiFe$_2$O$_4$ with either disorderly distributed or ordered Ni$^{2+}$ and Fe$^{3+}$ over the octahedral B-sites were obtained by means of a shell model (SM) using the General Utility Lattice Program (GULP).\cite{gulp}

In order to reduce the number of adjustable model parameters some approximations were applied. First, a valence shell was considered for the O atoms only while Ni and Fe were treated as rigid ions. Second, the van der Waals attractive interaction was considered to act only between O shells, while it was neglected for the Ni (Fe) core  -  O shell pairs. These assumptions are justified by the much higher polarizability of the O$^{-2}$ ion compared to that of the transition-metal ions. The rigid-ion approximation for transition-metals is a common approximation in the shell-model calculations on transition-metal oxides.\cite{Lewis}. The short-range interatomic interactions were modelled by a Buckingham potential: $V(r) = A \exp(-r/\rho) - C/r^6$, where a non-zero van der Waals constant $C$ was retained for the O shell - O shell pairs only.

The starting model parameters were taken from the widely utilized parameter set of Lewis and Catlow \cite{Lewis}, with a formal charge of +3 assigned to the Fe ions in both A- and B-positions. As a next step the Buckingham $A$ parameters for the Ni$^{+2}$ core - O shell  and Fe$^{+3}$ core - O shell pairs were optimized in order to reproduce the experimentally observed lattice parameters for NiFe$_2$O$_4$, the cubic face-centered lattice constant $a_c = 8.337~\text{\AA}$  and the fractional oxygen position $u = 0.831$. It was assumed at this stage that even if a cation ordering takes place in the B-positions, on a macroscopic scale (above the detection limit of the diffraction techniques) the material can be described in a cubic approximation. For this reason the fit was performed by setting equal partial occupancies of 0.5 for the Fe$^{+3}$ and Ni$^{+2}$ ions in the B-position of the ideal cubic $Fd\bar3m$ structure. This is equivalent to introduction of an ``average'' cation in B-position having charge, mass and short-range potential parameters, which are arithmetic means between those corresponding to Ni$^{+2}$ and Fe$^{+3}$. The as obtained shell-model parameters are summarized in
Table~\ref{Parameters}. Finally, the set of fitted parameters was used to calculate the $\Gamma$-point normal modes for the average-atom cubic $Fd \bar 3 m$ structure, which are assumed to mimic the positions of the main Raman bands in the case of complete cation disorder at the B-sites. The same parameters were utilized to optimize the lattice parameters of the ordered $\alpha$- and $\beta$-type structures and to calculate the corresponding $\Gamma$-point modes. The calculated normal mode frequencies for the three structural models are listed in Table~\ref{Frequencies}.

\subsubsection{Raman spectroscopy evidence for 1:1 ordering at the B-sites of NiFe$_2$O$_4$}
Unlike standard X-ray and neutron diffraction techniques, which are most sensitive to the long range order, the Raman scattering is more sensitive to the local short range order,  which may differ from the averaged long-range one. As discussed above, the 1:1 ordering at the B-sites gives rise to structures of tetragonal $P4_122$(\#91) or orthorhombic $Immb$(\#74) symmetry.

As it follows from the polarization selection rules, the fully symmetrical modes $A_{1g}$($Fd\bar{3}m$), $A_1$($P4_122$) or $A_g$($Imma$) can be identified by their stronger intensity in the $XX$ and $X'X'$ spectra compared to that in the $XY$ and $X'Y'$ spectra. Such are the Raman peaks at 140, 235, 381, 449, 487, 571, 593, and 704~cm$^{-1}$. Their number exceeds the expected single $A_{1g}$ mode for $Fd\bar{3}m$ or five $A_g$ modes for $Imma$ structure. Therefore, it is plausible to accept that at least part of these peaks originate from $\alpha$-type ordering at the octahedral sites, corresponding to local $P4_122$ structure. Indeed, these experimental frequencies show closest match with the following calculated frequencies of the $A_1$ modes in the $P4_122$ structure: 168, 253, 395, 498, 573, and 694~cm$^{-1}$. In the same time, the calculations show a large frequency gap between 387 and 605~cm$^{-1}$ in the $A_g$ channel of the $Imma$ structure, which makes the $\beta$-type ordering unlikely from a spectroscopic point of view.

\begin{figure}[t]
\includegraphics[width=8cm]{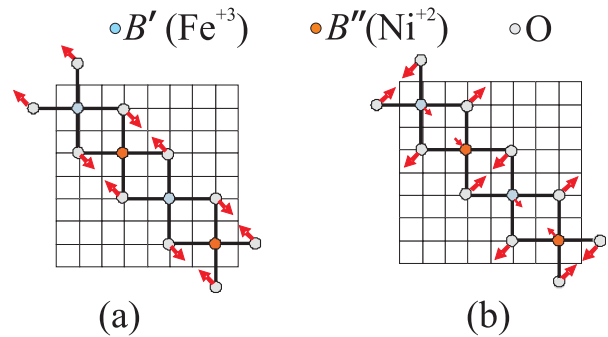}
 \caption{(Color online){ A $z=3/8$ cross-section of the $Fd\bar3m$ unit cell. Two FCC zone-boundary normal modes, which split into $A_1-B_2$ pairs upon $\alpha$-type 1:1 arrangement at $B$-positions: 571 - 593~cm$^{-1}$ (a) and 449 - 487~cm$^{-1}$ (b).}}
 \label{Folding}
\end{figure}

It is instructive to comment on the experimentally observed splitting of the intense band around ~ 580~cm$^{-1}$ into two components at 571 and 593~cm$^{-1}$. According to our calculations there are two closely separated modes in the $P4_122$ structure at 573 and 574~cm$^{-1}$ of $A_1$ and $B_2$ symmetry respectively. The selection rules for the tetragonal structure (see Table~\ref{Tetra}) predict the $B_2$ modes to appear in the same scattering configuration as $A_1$ modes. Thus the experimentally observed components can be assigned to an $A_1-B_2$ pair. The inspection of the atomic displacement pattern in these modes shows that they can be mapped to a doubly degenerated $X$-point zone-boundary mode of the cubic structure, which becomes a Raman-active $\Gamma$-point mode in the $P4_122$ structure due to zone folding (see Fig.~\ref{Folding}~(a)). Similarly, the 449 and 487~cm$^{-1}$ bands can be ascribed to another $A_1-B_2$ pair originating from the FCC zone-boundary $X$-point(see Fig.~\ref{Folding}~(b)). In a cubic structure the B$'$ and B$''$ sites are equivalent and the double degeneracy of each of these modes results from the fact that depending on the choice of the BO$_2$ chains the oxygen atoms can vibrate in two independent directions, e.g. $[110]_c$ and $[011]_c$, while the vibration in the third symmetry-equivalent direction $[101]_c$ is a linear superposition of the other two vibrations.

Additional pieces of evidence for the presence of tetragonal $\alpha$-type ordering could be drown from an analysis of the other scattering configurations. The mode at 213~cm$^{-1}$ is active in $X'X'$ and $XY$ configurations and could be assigned to $F_{2g}$, $B_1+E$, or $A_g+B_{2g}+B_{3g}$ vibrations in the cubic, tetragonal and orthorhombic structures respectively. Again, the best correspondence is found with the $E$-mode at 208~cm$^{-1}$ for the $P4_122$ structure. However, taking into account the unavoidable uncertainty of calculations, the $B_{2g}$ mode at 229~cm$^{-1}$ of the $Imma$ structure is also a like candidate for that spectral feature.

The cubic $E_g$ channel includes the $XX$, $X'X'$, and $X'Y'$ scattering configurations with a lowest expected intensity in the $X'X'$ configuration. Similar selection rules are expected for the $A_1+B_2$ modes in the $P4_122$ structure and the $A_g+B_{1g}$ modes in the the $Imma$ structure.  The experimentally detected modes at 273, 301, and 333~cm$^{-1}$ follow closely these selection rules. The multiplicity of the observed frequencies suggests the presence of ordered structures since one single $E_g$ mode is expected for the cubic structure. Most likely, the experimentally observed spectral bands correspond to the $A_1$ and $B_2$ modes of the $P4_122$ structure, whose calculated frequencies fall in the range 295 - 353~cm$^{-1}$ (see Table~\ref{Frequencies}). Since the frequency splitting between some of the modes (305-306 and 350-353~cm$^{-1}$) is comparable to the spectrometer resolution, experimentally they may appear as single spectral features leading to only three observable frequencies. It is worth mentioning that according to our calculations no $A_g$ or $B_{1g}$ modes are expected in this frequency range for the $Imma$ structure.

Finally, the band at 656-658~cm$^{-1}$ is active in all scattering configurations. According to Table~\ref{Ortho} such a behavior is expected for the $A_g$ modes of the orthorhombic $Imma$ structure. One plausible explanation of this feature is the calculated $A_g$ mode at 659~cm$^{-1}$. Alternatively, a doublet of closely separated modes of $B_2$ symmetry at 660.7~cm$^{-1}$ and of $E$ symmetry at 665.4~cm$^{-1}$ is predicted for the tetragonal $P4_122$ structure. Due to their similar frequencies the two modes may be indistinguishable experimentally, and should appear in all scattering configuration ( Table~\ref{Tetra}).

\section{Summary and Conclusions}
We present detailed polarization Raman measurements of the inverse spinel NiFe$_2$O$_4$. The number and the polarization selection rules for the observed Raman bands do not support the model of stochastic distribution of Ni$^{2+}$ and Fe$^{3+}$ cations among the octahedral $B$-sites.

By using symmetry analisys and shell-model lattice dynamics calculations we examined the experimental data against two models of $B$-site 1:1 ordering, $\alpha$-type of tetragonal $P4_122/P4_322$ symmetry and $\beta$-type of orthorhombic $Imma$ symmetry. All experimental Raman bands are consistent by symmetry and frequency with the calculated normal modes of the $\alpha$-type structure, The $\beta$-type ordering can explain only part of the observed bands.

On the basis of the above arguments we can conclude that Ni$^{2+}$ and Fe$^{3+}$ exibit 1:1 ordering at the octahedral sites of NiFe$_2$O$_4$, most probably of tetragonal $P4_122/P4_322$ symmetry. However, the $Imma$ structure can not be definitively ruled out due to the good correspondence of some calculated frequencies to experimentally observed bands. It is possible that the two types of ordering coexist with prevalence of the tetragonal structures. Macroscopically the material exhibits cubic symmetry due to the presence of randomly oriented twin variants of the ordered structures.

\acknowledgements
This work is supported in part by the State of Texas through The Texas Center for Superconductivity at the University of Houston and partly by the contracts \# DO 02-167/2008 and TK-X-1712/2007 of the Bulgarian National Scientific Research Fund.

\newpage

\begin{table}
\caption []{Raman polarization selection rules for the spinel $Fd\bar{3}m$ structure.}
\label{CubicSR}
\

\begin{tabular}{|c |c| c |c |c |}
\hline
Mode        & $XX$      &$X'X'$    & $XY$  & $X'Y'$ \\
\hline
$A_{1g}$    &$a^2$      &$a^2$     & 0     & 0 \\
$E_g$       &$4b^2$     &$b^2$     & 0     & $3b^2$ \\
$F_{2g}$    &  0        &$d^2$     & $d^2$ &  0\\
\hline

\end{tabular}
\end{table}

\begin{table*}

\caption []{Atomic site symmetries and corresponding $\Gamma$-point modes for the $Fd\bar{3}m$,  $P4_122$, and $Imma$ structures of NiFe$_2$O$_4$. For the ordered $P4_122$ and $Imma$ structures only Raman modes are shown.}
\label{Modes}
\

\begin{tabular}{|c c c |c c c|c c c|}
\hline
&  &  &  &  &  &  &  &  \\
\multicolumn{3}{|c|}{$Fd\bar{3}m$ (cubic)} &  \multicolumn{3}{|c|}{$P4_122$ (tetragonal)} &

\multicolumn{3}{|c|}{$Imma$ (orthorhombic)} \\
&  &  &  &  &  &  &  &  \\
\hline
&  &  &  &  &  &  &  &  \\
       &Wickoff  &$\Gamma$-point  &       & Wickoff & Raman &      & Wickoff &Raman  \\
  Atom &index    &modes  & Atom  & index   & modes &Atom   & index   & modes \\
  &  &  &  &  &  &  &  &  \\
        \hline
   &  &  &  &  &  &  &  &  \\
 Fe(1)    &8a  &$F_{2g}+F_{1u}$  & Fe(1)  &4c  & $A_1+2B_1+B_2+3E$ & Fe(1) & 4e &

$A_g+B_{2g}+B_{3g}$  \\
          &  &  &  &  &  &  &  &  \\
 Ni/Fe(2) & 16d & $A_{2u}+E_{u}+F_{2u}+2F_{1u}$& Ni     & 4a & $A_1+B_1+2B_2+3E$ & Ni    &

4c &  ---\\
          &     &         & Fe(2)  & 4b & $A_1+B_1+2B_2+3E$ & Fe(2) & 4b &  ---\\
          &  &  &  &  &  &  &  &  \\
O  & 32e & $A_{1g}+A_{2u}+E_g+E_u+$ & O(1) & 8d &$3A_1+3B_1+3B_2+6E$&O(1)&8h

&$2A_g+B_{1g}+B_{2g}+2B_{3g}$\\
   &     & $F_{2u}+2F_{2g}+2F_{1u}+F_{1g}$ & O(2) & 8d &$3A_1+3B_1+3B_2+6E$&O(2)&8i

&$2A_g+B_{1g}+2B_{2g}+B_{3g}$\\
          &  &  &  &  &  &  &  &  \\
          \hline
            &  &  &  &  &  &  &  &  \\
\multicolumn{2}{|l}{TOTAL Raman:}& $A_{1g}+E_g+3F_{2g}$ &  \multicolumn{3}{|c|}{TOTAL:\ \ \

$9A_1+10B_1+11B_2+21E$ } & \multicolumn{3}{|c|}{TOTAL: $5A_g+2B_{1g}+4B_{2g}+4B_{3g}$} \\
\multicolumn{2}{|l}{1 acoustic + 4 IR:}& $5F_{1u}$ &  \multicolumn{3}{|c|}{  } &

\multicolumn{3}{|c|}{} \\
\multicolumn{2}{|l}{inactive:}&$2A_{2u}+2E_u+2F_{2u}+F_{1g}$ &  \multicolumn{3}{|c|}{  } &

\multicolumn{3}{|c|}{} \\
&  &  &  &  &  &  &  &  \\
\hline

\end{tabular}
\end{table*}

\begin{table*}

\caption []{Polarization selection rules for the scattering from $P4_122$ structures averaged over the three twin variants with I-,II,- and III-type orientation with respect to the cubic axes.}
\label{Tetra}
\

\begin{tabular}{|c|c|c|c|c|c|}
\hline
 &  &  &  &  & \\
Mode &Raman tensor&$XX$ (cubic)&$XY$ (cubic)&$X'X'$ (cubic)&$X'Y'$ (cubic)\\
 &  &  &  &  & \\

\hline

 &  &  &  &  & \\

$A_1$&  $\left[ \begin{array}{ccc}
a &  &  \\
 & a &  \\
 &  & b \end{array} \right] $ &$\frac{2}{3}a^2+\frac{1}{3}b^2$& 0

&$\frac{1}{3}a^2+\frac{1}{6}(a+b)^2$&$\frac{1}{6}(a-b)^2$\\

 &  &  &  &  & \\

 \hline

  &  &  &  &  & \\

$B_1$&  $\left[ \begin{array}{ccc}
c &  &  \\
 & -c &  \\
 &  &  \end{array} \right] $  &0& $\frac{1}{3}c^2$ &$\frac{1}{3}c^2$&0\\

 &  &  &  &  & \\
\hline
 &  &  &  &  & \\

$B_2$&  $\left[ \begin{array}{ccc}
 & d &  \\
d &  &  \\
 &  &  \end{array} \right] $  &$\frac{2}{3}d^2$& 0 &$\frac{1}{6}d^2$&$ \frac{1}{2}d^2$
 \\

 &  &  &  &  & \\
\hline
 &  &  &  &  & \\

  &  $\left[ \begin{array}{ccc}
\  &\   &\   \\
\  &\   & e \\
\  & e & \  \end{array} \right] $  & & & &   \\

$E$&  &  $0$& $\frac{2}{3}e^2$ &$\frac{2}{3}e^2$&$0$\\

&  $\left[ \begin{array}{ccc}
\  & \  &-e  \\
\  & \  & \  \\
-e & \  & \  \end{array} \right] $  & & & &   \\

 &  &  &  &  & \\
\hline

\end{tabular}
\end{table*}

\begin{table*}

\caption []{Polarization selection rules for the scattering from $Imma$ structures averaged over the six twin variants with I(IV)-,II(V),- and III(VI)-type orientation with respect to the cubic axes.}
\label{Ortho}

\

\begin{tabular}{|c|c|c|c|c|c|}
\hline
 &  &  &  &  & \\
Mode & Tensor &$XX$ (cubic)&$XY$ (cubic)&$X'X'$ (cubic)&$X'Y'$ (cubic)\\
 &  &  &  &  & \\
\hline

\

 &  &  &  &  & \\

$A_1$&  $\left[ \begin{array}{ccc}
a &  &  \\
 & b &  \\
 &  & c \end{array} \right] $

&$\frac{1}{6}(a+b)^2+\frac{1}{3}c^2$&$\frac{1}{12}(a-b)^2$&$\frac{1}{6}a^2+\frac{1}{6}b^2+\frac{1}{24}(a+b+2c)^2$&$\frac{1}{24}(a+b-2c)^2$\\

 &  &  &  &  & \\
\hline
 &  &  &  &  & \\

$B_{1g}$&  $\left[ \begin{array}{ccc}
 & d &  \\
d &  &  \\
 &  &  \end{array} \right] $  &$\frac{2}{3}d^2$& {0} &$\frac{1}{6}d^2$&$\frac{1}{2}d^2$\\

 &  &  &  &  & \\
\hline
 &  &  &  &  & \\

$B_{2g}$&  $\left[ \begin{array}{ccc}
 &  & e \\
 &  &  \\
 e&  &  \end{array} \right] $  & $\frac{1}{3}e^2$ &$\frac{1}{3}e^2$\\

 &  &  &  &  & \\
\hline
 &  &  &  &  & \\

$B_{3g}$&  $\left[ \begin{array}{ccc}
&  &   \\
  &   & f \\
  & f &   \end{array} \right] $  &$0$& $\frac{1}{3}f^2$ &$\frac{1}{3}f^2$&$0$\\

 &  &  &  &  & \\
\hline

\end{tabular}
\end{table*}

\begin{table}
\caption{Shell-model parameters for NiFe$_2$O$_4$.}

\label{Parameters}

\

\begin{tabular}{|c|c|c|c|}
\hline
 & & & Core-shell\\
Atom & Core charge & Shell charge &  spring constant  \\
 & $X$  & $Y$  & $k$~(eV/\AA$^2$) \\
 \hline
 Ni & +2 & --- & --- \\
 Fe & +3 & --- & --- \\
  O & 0.513 & -2.513 & 72.53 \\
\hline
\hline
 & & & \\
Atomic pair & $A$ (eV) & $\rho$ (\AA) & $C$~(eV $\times$ \AA$^6$) \\
 & & & \\
 \hline
 Ni core - O shell & 681.9 & 0.337 & 0 \\
 Fe core - O shell & 986.1 & 0.337 & 0 \\
 O shell - O shell & 22764.0 & 0.149 & 27.879 \\
 \hline
\end{tabular}

\end{table}

\begin{table}

\caption []{Calculated frequencies of the Raman-active modes in NiFe$_2$O$_4$ for the three structural models: $B$-site disorder of a macroscopic $Fd \bar3 m$ symmetry, $\alpha$-type ordering (symmetry $P4_1 22$) and $\beta$-type ordering (symmetry $Imma$)}.
\label{Frequencies}

\begin{tabular}{|c|c|c|c|c|c|c|c|c|c|c|}
\hline
\multicolumn{3}{|c|}{} & \multicolumn{4}{c|}{} & \multicolumn{4}{c|}{} \\
\multicolumn{3}{|c|}{$Fd \bar3 m$} & \multicolumn{4}{c|}{$P4_1 22$} & \multicolumn{4}{c|}{$Imma$} \\
\multicolumn{3}{|c|}{} & \multicolumn{4}{c|}{} & \multicolumn{4}{c|}{} \\
\hline
$A_g$ & $E_g$ & $F_{2g}$ & $A_1$ & $B_1$ & $B_2$ & $E$ & $A_g$ & $B_{1g}$ & $B_{2g}$ & $B_{3g}$ \\
\hline
 &     &     & 168 & 148 & 155 & 147 &  &  &  &  \\
 &     &     &     & 171 & 205 & 207 &  &  &  &  \\
 &     &     &     &     & 235 & 235 & 227 &  &  & 229 \\
 &     & 255 & 253 & 248 &     & 246 &  & 248 &  &  \\
 &     &     &     &     &     & 253 &  &  & 252 &  \\
 &     &     &     &     &     & 263 &  &  &  &  \\
 & 300 &     & 295 &     & 305 & 293 &  &  &  &  \\
 &     &     &     & 329 & 306 & 314 &  &  &  &  \\
 &     &     & 350 &     & 353 & 328 &  &  & 336 &  \\
 &     &     & 395 & 381 & 374 & 367 & 387 &  &  &  \\
 &     &     &     & 415 &     & 407 &  &  &  & 405 \\
 &     &     &     &     &     & 438 &  & 438 &  &  \\
 &     & 465 &     & 465 &     & 464 &  &  &  &  \\
 &     &     &     &     &     & 469 &  &  &  &  \\
 &     &     & 498 & 498 & 496 & 490 &  &  &  &  \\
 &     &     &     &     & 516 & 514 &  &  &  &  \\
 &     &     & 573 &     & 574 & 556 &  &  & 588 &  \\
 &     & 618 &     & 612 &     & 615 & 605 &  &  & 613 \\
 &     &     & 635 & 649 & 661 & 665 & 659 &  & 643 &  \\
687 &  &     & 694 &     &     & 718 &  &  &  & 689 \\
 &     &     &     &     &     &     & 744 &  &  &  \\
\hline

\end{tabular}
\end{table}

\end{document}